\documentclass[aps,prb,twocolumn,superscriptaddress]{revtex4-2}
\usepackage{graphicx,epstopdf,siunitx,braket,xspace,amssymb,dsfont,bm,amsmath,float}
\usepackage{comment}
\usepackage[linktocpage=true,colorlinks=true,urlcolor=blue,linkcolor=red,citecolor = blue]{hyperref}
\usepackage{pgfplots}
\usepackage{amsfonts}
\usepackage{xcolor}

\newcommand{\DU}[1]{\ensuremath{\ket{\downarrow \uparrow}}}

\newcommand{\sige}[0]{Si$_{0.7}$Ge$_{0.3}$\xspace} 
\newcommand{\HfOx}[0]{HfO$_2$\xspace}
\newcommand{\SiOx}[0]{SiO$_2$\xspace}
\newcommand{\Vtop}[0]{$V_{\mathrm{top}}$\xspace}

\begin{document}

\title{On-chip cryogenic multiplexing of Si/SiGe quantum devices}

\author{M. A. Wolfe}
\affiliation{Department of Physics, University of Wisconsin-Madison, Madison, WI 53706, USA}

\author{Thomas McJunkin}
\thanks{Present address: John Hopkins University Applied Physics Laboratory, Laurel, Maryland 20723}
\affiliation{Department of Physics, University of Wisconsin-Madison, Madison, WI 53706, USA}

\author{Daniel R. Ward}
\thanks{Present address: HRL Laboratories, LLC, 3011 Malibu Canyon Road, Malibu CA 90265, USA}
\affiliation{Sandia National Laboratories, Albuquerque, NM 87185, USA.}

\author{DeAnna~Campbell}
\affiliation{Sandia National Laboratories, Albuquerque, NM 87185, USA.}

\author{Mark Friesen}
\affiliation{Department of Physics, University of Wisconsin-Madison, Madison, WI 53706, USA}

\author{M. A. Eriksson}
\affiliation{Department of Physics, University of Wisconsin-Madison, Madison, WI 53706, USA}

\begin{abstract}
The challenges of operating qubits in a cryogenic environment point to a looming bottleneck for large-scale quantum processors, limited by the number of input-output connections. 
Classical processors solve this problem via multiplexing; however, on-chip multiplexing circuits have not been shown to have similar benefits for cryogenic quantum devices.
In this work we integrate classical circuitry and Si/SiGe quantum devices on the same chip, providing a test bed for qubit scale-up. 
Our method uses on-chip field-effect transistors (FETs) to multiplex a grid of work zones, achieving a nearly tenfold reduction in control wiring. 
We leverage this set-up to probe device properties across a 6$\times$6~mm$^2$ array of 16 Hall bars.
We successfully operate the array at cryogenic temperatures and high magnetic fields where the quantum Hall effect is observed. 
Building upon these results, we propose a vision for readout in a large-scale silicon quantum processor with a limited number of control connections.

\end{abstract}
	
\maketitle

\section{Introduction}

Quantum processors in silicon present opportunities for integrating quantum bits (qubits) with classical on-chip control logic, due to the materials compatibility of these two types of circuits~\cite{Vandersypen2017,  Boter2019, Franke2019, Gonzalez-Zalba2021}. 
Fault-tolerant operation of these devices requires one- and two-qubit gate fidelities in excess of 99\%, as recently demonstrated in state-of-the-art quantum processors in silicon~\cite{Dehollain2016, Yoneda2018, Huang2019, Petit2020,Xue2022,Mills2022_2}.
However, fault tolerance also requires very large numbers of qubits~\cite{Fowler2012}, which is at odds with cryogenic considerations that constrain the number of wires leading from room-temperature electronics to the quantum processor.
In current Si qubit implementations, the number of input-output (I/O) connections for 
quantum processors is proportional to the number of qubits, which fundamentally limits the qubit count in a quantum processor.
This obstacle is known as the interconnect bottleneck, and it poses a serious challenge for scale-up. 

Classical microprocessors address the interconnect bottleneck by employing well-developed multiplexing technologies to pack billions of transistors into a 4$\times$4~cm$^2$ package, with only $\sim 10^{3}$ I/O connections~\cite{intel}.
The relationship between the number of transistors, $g$, to I/O connections, $T$, takes the form of a power-law, $T\propto g^p$, where $p$ is the known as the Rent exponent.
For devices with no intentional reduction of interconnects, we naturally have $p\approx 1$.
Modern classical processors, in contrast, have been engineered to achieve low Rent exponents, $p = 0.36$~\cite{Davis1998}. 
For quantum processors, an error-corrected implementation of Shor's algorithm involves $\sim 10^9$ physical qubits~\cite{Fowler2012} and could therefore benefit from a similarly small Rent exponent.
Solutions to this problem could involve technologies such as cryogenic signal generation~\cite{Bardin2019,VanDIjk2020,Patra2020, Pauka2021,Xue2021_1}, cryogenic readout  
hardware~\cite{Homulle2016,Bagas2021,Guevel2021, Ruffino2022}, cross-bar network architectures~\cite{Veldhorst2017,Li2018}, and signal multiplexing~\cite{Ward2013,Al-Taie2013,Puddy2015,Schaal2019, PaqueletWuetz2020,Bohuslavskyi2022, Thomas2023, Acharya2023}. 
Multiplexing, which is the topic of this work, is especially effective at addressing large arrays of devices, such as transistors or Si/SiGe quantum dots, using integrated field-effect transistor (FET) switches to control a common bus~\cite{Ward2013}. 
As we show below, this provides a natural scheme for performing readout in a large quantum processor.

In this letter, we experimentally demonstrate the integration on a single chip of a classical multiplexing circuit with Si/SiGe quantum devices. In our implementation, the multiplexing circuit addresses 16 construction zones with an exponential reduction in external control wires.
Moreover, the scheme is suitable for use in a wide range of gate-defined quantum devices. 
To demonstrate the power of the technique, we use electron-beam lithography to fabricate Hall bars in each of the construction zones, which we use to assess device uniformity across the whole chip in a single cool-down. The design integrates FETs to electrically switch between the Hall bars, which make up an array of sensors capable of probing the heterostructure quality. We report diagnostics relevant for quantum devices in silicon, such as threshold voltage, Hall bar capacitance, percolation density, electron mobility, and defect density. Our results indicate notable uniformity in the electron mobility (relative standard deviation $\approx$ 10\%) and percolation density ($\approx$ 15\%). The quantum and classical circuits are operated at a temperature of $\sim$\SI{2}{K}, similar to recent realizations of hot silicon qubits~\cite{Yang2020,Petit2020}, using only a pumped $^4$He cooling circuit, which provides more cooling power than $^3$He systems and is a strong candidate for use with large-scale quantum processors. We observe the integer quantum Hall effect in the device array, even at these elevated temperatures. We discuss the operation of the on-chip multiplexer (MUX) at high magnetic fields, and we describe a protocol for multiplexed readout of quantum-dot qubits and discuss the operating requirements of the multiplexing circuitry for high-fidelity readout. Our work demonstrates a test bed for high-throughout device characterization, and provides a framework for silicon quantum processors with integrated on-chip classical circuitry. 

\section{Integration of quantum and classical circuits}

\begin{figure}
    \includegraphics[width = \columnwidth]{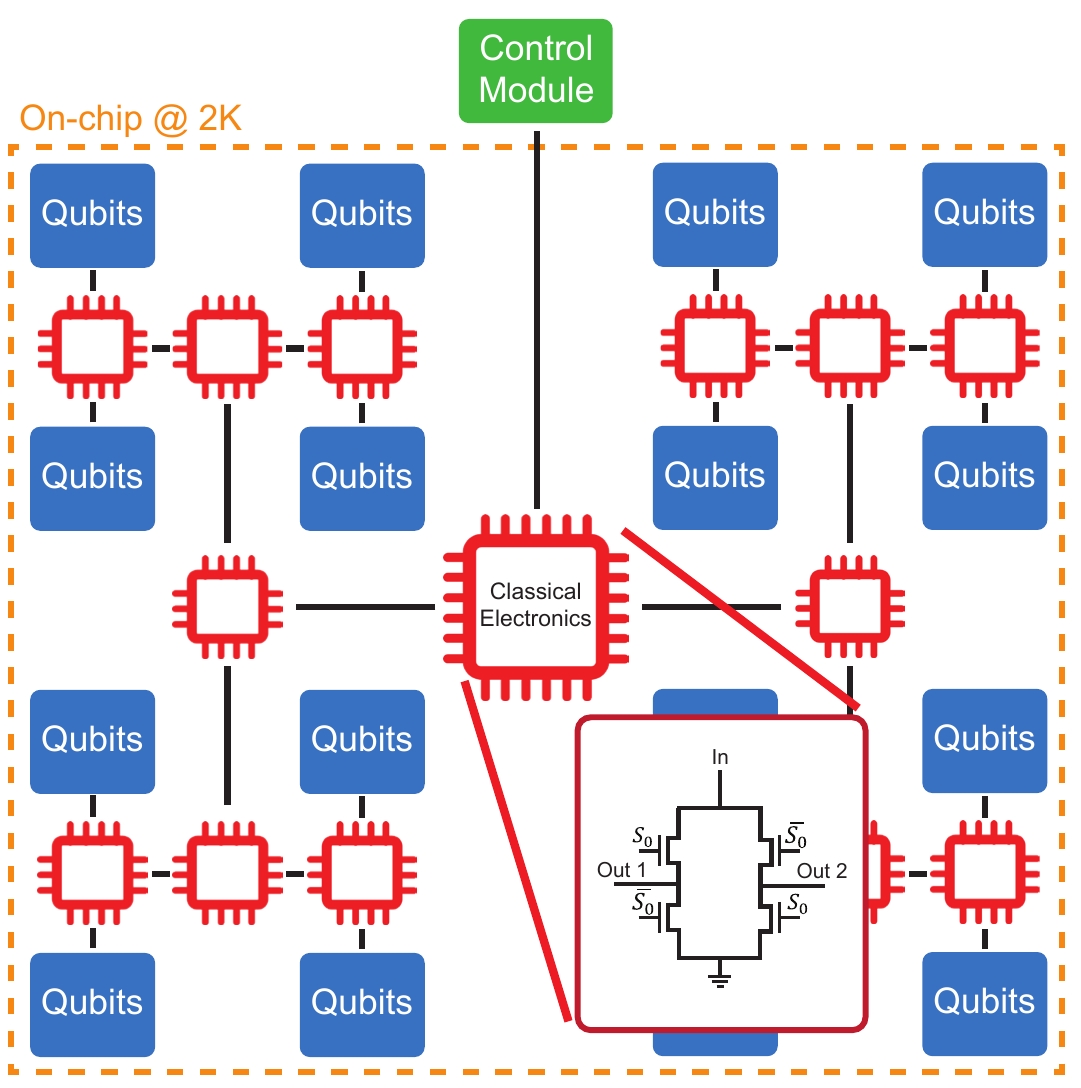}
    \caption{Blueprint for a quantum processor with on-chip, integrated classical circuitry. Blue boxes indicate dense qubit arrays that require input and output signals. Signals are generated and processed by a shared control module (green box), and routed via an array of on-chip switches (red boxes). The circuit for the first switch (inset) is controlled by the first bit $S_0$ of the binary address $S$, with complement $\bar S$.}
    \label{fig1}
\end{figure}

We first describe a simple layout for dispersing integrated control electronics between tiled qubit arrays, as shown in Fig.~\ref{fig1}. Inspired by Vandersypen et al.~\cite{Vandersypen2017}, who proposed a layer of classical electronics occupying the space next to qubits, our design positions on-chip classical circuits (red boxes) between sparse qubit arrays (blue boxes), all patterned on the same chip. This layout focuses on providing each qubit array with necessary I/O signals by routing to a shared control module (green box), which could be designed on-chip or off-chip. This module can host a variety of cryogenic electrical hardware, such as cryoCMOS controllers, to enable qubit control and readout~\cite{Gonzalez-Zalba2021}. The key function of this layout is to steer the control signals to the appropriate qubit array using on-chip classical electronics.

Below, we demonstrate on-chip multiplexing, enabled by a grid of switches, each comprised of four FETs, as illustrated in the inset of Fig.~\ref{fig1}. When address $S$ is sent to the multiplexer, the FETs connect a specific qubit array to the control module, while the other arrays remain idle. As elaborated in the following section, this is accomplished by the switches splitting the signal into two possible paths, each in turn doubling the number of addressable arrays for each added switch. Equipping a quantum chip with many such switches allows the I/O controls for a single qubit array to be shared across an entire grid of arrays, each with a unique address. This design allows for far fewer connections between the control module and chip, thus significantly reducing the Rent exponent of the quantum processor. 
Indeed, in this work we achieve an exponential reduction of I/O connections used for readout.

\section{Multiplexing Scheme}

\begin{figure*} 
    \includegraphics[width = 2.1\columnwidth]{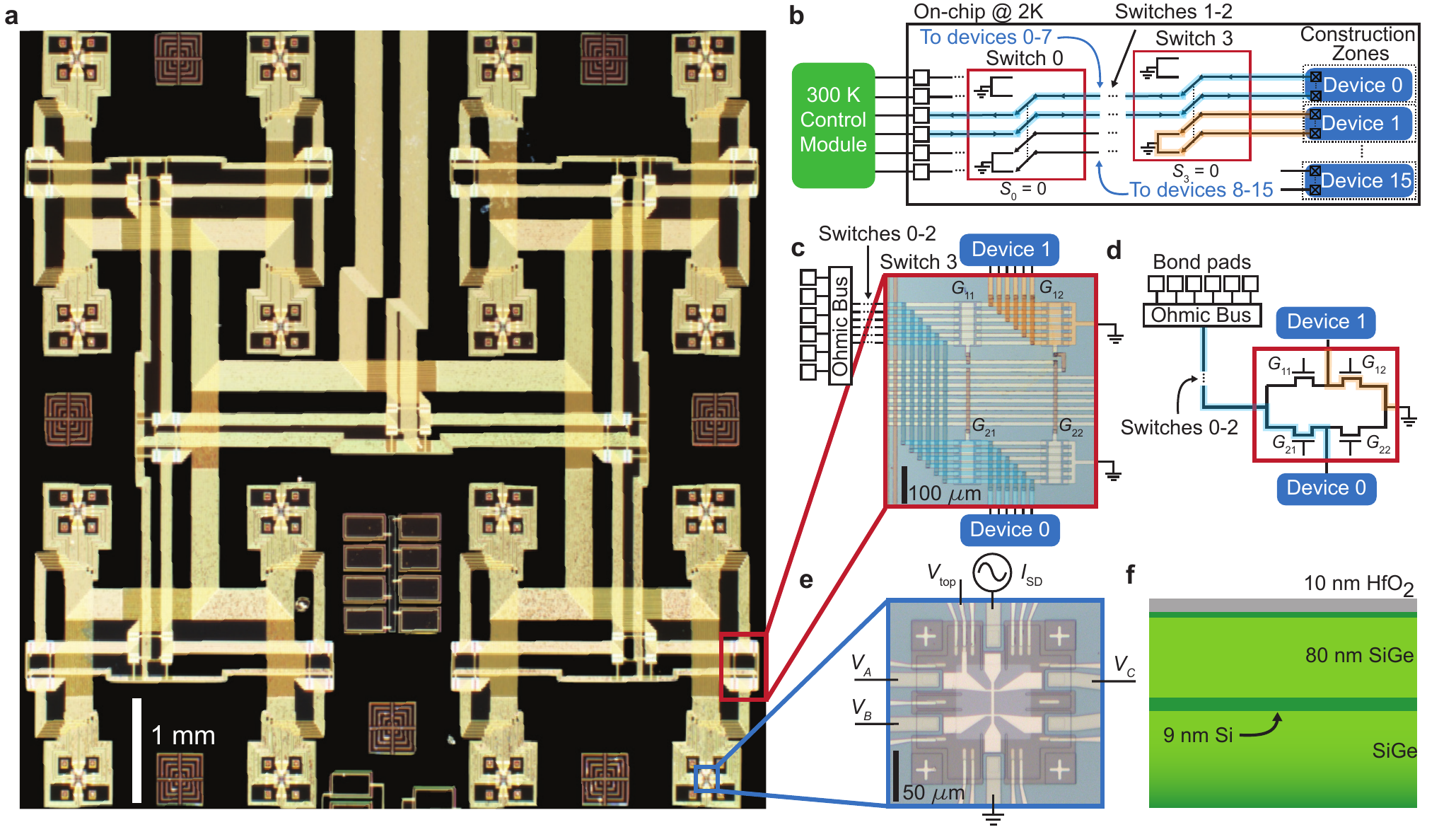}
    \caption{Overview of multiplexing scheme. 
    (a) Chip design featuring 16 construction zones. (This image has been stitched and its background has been removed for clarity. See Supplementary Sec.~S1 for the image without the removed background and CAD design.) 
    (b) Circuit schematic illustrating the device connections to the \SI{300}{K} control module via the switching network. 
    (Not all lines are shown, for brevity.)
    Lines highlighted in blue show current routed to and from Device~0, while lines to other devices are grounded (orange).
    (c) Micrograph image of Switch~3 showing corresponding false-color, highlighted current paths, for the digital state $S_3 = 0$. 
    (d) FET circuit schematic for Switch~3 with corresponding highlighted current paths. 
    (e) Micrograph of the construction zone at address $S = 0000$ hosting Device~0 (a Hall bar). 
    Key elements of the AC lock-in measurement scheme are labeled. 
    (f) Heterostructure stack of the mesa located at the center of each construction zone.}
    \label{fig2}
\end{figure*}  

Figure~\ref{fig2}(a) shows the device used in our experiment, comprised of sixteen construction zones fabricated on a single 11.5$\times$11.5~mm$^2$ chip containing a Si/SiGe quantum well.
Each of these zones provides a platform for implementing a quantum circuit in a sparse array. 
At the center of each zone is an undoped heterostructure mesa containing six ion-implanted ohmic contacts, alignment markers, palladium jumpers, and a terraced gate oxide. 
More details are given in Methods.
This structure provides the ingredients for constructing a variety of lithographically defined quantum devices, including quantum-dot qubits~\cite{Dodson2020}. 
In our experiments, we form Hall bars in the constructions zones, as described in the following section. 
 
To address the device zones individually with a multiplexer, we assign each zone a unique binary digital address $S$.
For the 16-zone device considered here, we need a four-bit address: $S = \{S_0, S_1, S_2, S_3\}$, where $S_i$ denotes the digital state of Switch~$i$.
In the current work, only the ohmic contacts are multiplexed, enabling control of the current paths for measurement running through the two-dimensional electron gas (2DEG) to each of the device zones. 
In Fig.~\ref{fig2}(b), the switched current path leading from the room-temperature electronics to Device~0 with address $S=0000$ is indicated in blue.  
The return current path is also switched, as also indicated in blue. 
We note that previous implementations of multiplexed quantum devices in silicon left the inactive devices floating~\cite{Ward2013}, which caused the drains of the FETs in those circuits to charge up to an unknown voltage, resulting in voltage drift and uncontrolled hysteresis~\cite{Ward_private}. 
We solve this problem here by introducing additional FETs that provide ground paths to the inactive devices.
For example, the inactive Devices~1-15 in Fig.~\ref{fig2}(b) are all grounded, with an example current path for Device~1 indicated in orange.

The circuit for a single switch is shown in Fig.~\ref{fig2}(d), with four FETs labeled $G_{ij}$, and the active (blue) and grounded (orange) current paths also indicated. 
Here, pairs of FETs are tied together such that $\{G_{11},G_{22}\}$ have the same input and $\{G_{12},G_{21}\}$ have the opposite input; in this way, the switch can take only two digital states. 
For the digital state corresponding to $S_3=0$ in this example, $V_{G_{12}} = V_{G_{21}}$ is positive and $V_{G_{11}} = V_{G_{22}}$ is negative, and vice-versa for $S_3=1$.  
The micrograph image in Fig.~\ref{fig2}(c) shows the full set of FETs comprising Switch~3.
Here again, the false coloring shows the active (blue) and grounded (orange) current paths. 
For the Hall bar devices described in this work, a total of six ohmic contacts are multiplexed, corresponding to the six switched lines shown in Fig.~\ref{fig2}(c).

\section{Transport and Magnetoresistance Statistics}

\begin{figure*}
    \includegraphics[width = 2.1\columnwidth]{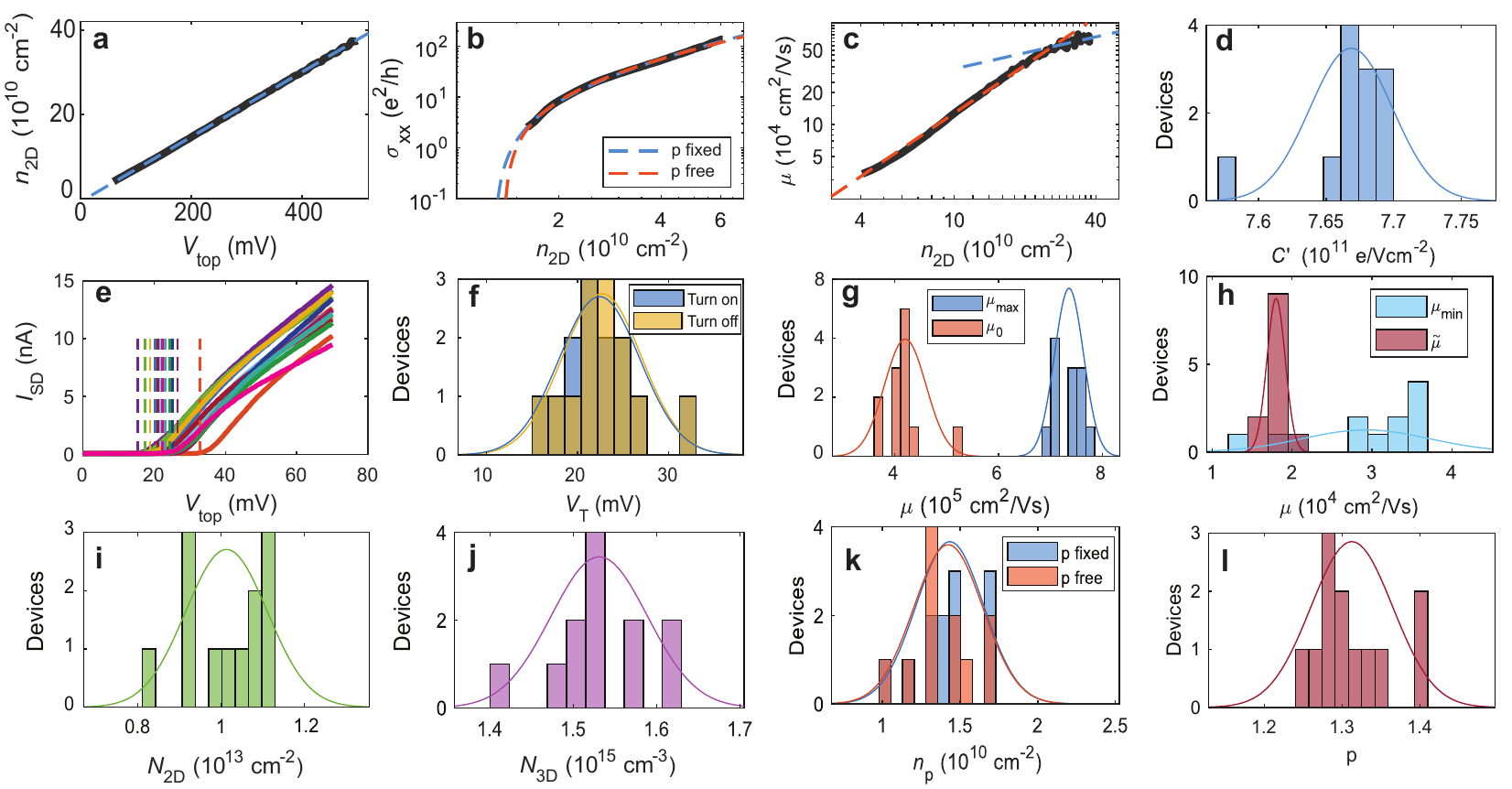}
    \caption{Multiplexed device properties. 
    (a) Capacitive charging curve for the $S = 0000$ device. The blue dashed line is the linear best-fit, with a slope defined as $C'$.
    (b) Longitudinal conductivity $\sigma_{xx}$ vs $n_{\rm{2D}}$, for the same device, showing percolation conductance at low densities. The blue dashed line shows the theoretical fit when with the percolation exponent $p$ is allowed to be a fitting parameter, the red dashed line shows the result when we set $p = 4/3$.
    (c) Mobility vs density for the same device, showing fits to different power laws: $n_{\rm{2D}}^{3/2}$ (blue) vs $n_{\rm{2D}}^{1/2}$ (red), where the former describes 2D scattering from remote charge defects, while the latter describes 3D scattering from uniform background charges in the quantum well. 
    (d) Statistics of the Hall bar capacitance.
    (e) Turn-on curves for the measured Hall bars. 
    Vertical dashed lines indicate the corresponding threshold voltages. 
    (f) Statistics of the threshold voltages.
    (g) Statistics of the mobility obtained at a fixed value of the density $n_{\rm{2D}} = \SI{2e11}{cm^{-2}}$ ($\mu_0$) vs the maximum mobility ($\mu_{\rm{max}}$). 
    (h) Statistics of lowest measured mobility ($\mu_{\rm{min}}$) vs the mobility estimated from turn-on curves ($\tilde{\mu}$). 
    (i) Statistics of the theoretically determined defect densities in the oxide, $N_{\rm{2D}}$.
    (j) Statistics of the theoretically determined defect densities in the quantum well, $N_{\rm{3D}}$.
    (k) Statistics of the percolation density when the percolation exponent is allowed to be a fitting parameter (blue), or when it is set to the theoretically predicted value $p=1.31$. 
    (p) Statistics of the percolation exponent $p$. }
    \label{fig3}
\end{figure*}

We now make use of on-chip multiplexing to characterize device uniformity across an array of Si/SiGe devices. 
In these experiments, a single Hall bar is fabricated atop each of the 16 construction zones.
The Hall bar layout is shown in Fig.~\ref{fig2}(e), with the top-gate voltage labeled $V_\text{top}$, voltage probe contacts labeled $V_A$, $V_B$, and $V_C$, source-drain current source labeled $I_\text{SD}$, and the Hall bar ground as indicated. (See Methods for further details.) Four-probe AC lock-in measurements are used to determine the magnetoresistance, where the longitudinal voltage $V_{xx} = V_A - V_B$ and transverse Hall voltage $V_{xy} = V_A - V_C$ are measured simultaneously. 
The elements of the resistivity tensor, $\rho_{xx} = aV_{xx}/I_{\text{SD}}$ and $\rho_{xy} = V_{xy}/I_{\text{SD}}$, characterize the density of the two-dimensional electron gas (2DEG) through the relation $n_\text{2D} = B/(e\rho_{xy})$, where $B$ is the magnetic field applied out of plane, and $e$ is the (positive) electron charge.
Hall bar measurements are also used to characterize the zero-field longitudinal conductance $\sigma_{xx} = 1/\rho_{xx}$ and the mobility $\mu = 1/(en_{\rm{2D}}\rho_{xx})$ in the classical (low-field, $B < \SI{0.5}{T}$) Hall regime. 
Here, we set $I_{\rm{SD}} = \SI{10}{nA}$, and note the geometric aspect ratio of the Hall bar given by $a=1/8$. 
Some typical Hall scans are shown in Figs.~\ref{fig3}(a)-\ref{fig3}(c) for the device labeled $S=0000$.
Note that in Figs.~\ref{fig3}(b) and \ref{fig3}(c), the quantities $\sigma_{xx}$ and $\mu$ are plotted as implicit functions of $n_\text{2D}(V_\text{top})$.
Below, we perform similar scans on the other Hall bars and collect statistics on their operation.

We first explore the uniformity of the geometric and dielectric properties of the Hall bars.
A simple check of device uniformity is given by the Hall bar capacitance per unit area $C'$.
As consistent with data shown in Fig.~\ref{fig3}(a), we consider scans obtained in the linear regime, above the turn-on voltage (described below), on which we perform a least-squares fit to the formula $C'=n_\text{2D}/V_\text{top}+c_0$, which includes a small offset $c_0$.
The resulting histograms are shown in Fig.~\ref{fig3}(d).
The data exhibit a very narrow range of capacitance values with a mean value of $\bar{C'} = \SI{7.67e11}{e/Vcm^2}$ and a relative variance of 0.4\%, indicating a high-quality gate stack. 

A second device property of interest is the turn-on voltage $V_\text{T}$, defined as the value of $V_\text{top}$ needed to achieve a current flow of $I_\text{SD}=\SI{130}{pA}$.
Turn-on curves for all the measured devices are shown in Fig.~\ref{fig3}(e). 
Here, the turn-on voltages are indicated by vertical dashed lines, and their corresponding histogram is shown in Fig.~\ref{fig3}(f) (blue data).
Noting the presence of a small hysteresis between ramp-up and ramp-down, we can similarly define a turn-off voltage, with results also shown in Fig.~\ref{fig3}(f) (orange data).
The average threshold voltage in these measurements is quite low, $\bar V_\text{T}=\SI{22.2}{mV}$, indicating a relatively small amount of trapped charge at the oxide interface. 
The narrow spread in threshold voltages, $\Delta V_\text{T}=\SI{4}{mV}$, indicates geometrical uniformity of the Hall bars and consistency of trapped charge beneath the various gates. 
We note that the FET switches in the multiplexer also all turn on in a similar voltage range; however, we normally operate those FETs well above their threshold voltages to mitigate unwanted Hall effects at higher magnetic fields. 
(See Supplementary Sec.~SIV for details.)

Electron mobility is a particularly important indicator of device performance.
Below, we characterize the mobility in several different ways.
Figure~\ref{fig3}(c) shows a typical scan of mobility vs 2DEG density, based on the standard definition $\mu = 1/(en_{\rm{2D}}\rho_{xx})$.
Figure~\ref{fig3}(g) shows a histogram of the maximum mobility values $\mu_\text{max}$ obtained for each of the devices (blue data).
The results are all quite high, with an average value of $\SI{695080}{cm^2/Vs}$ and a maximum value greater than \SI{780000}{cm^2/Vs}.
It is also useful to compare mobilities obtained at a single, fixed density for every device, ensuring an equivalent screening response from each of the accumulated 2DEGs. 
Here we define $\mu_0$ as the mobility obtained at the density $n_{\rm{2D}} = \SI{2e11}{cm^{-2}}$.
The resulting histogram of $\mu_0$ values is also shown in Fig.~\ref{fig3}(g) (red data).
In this case, the data have a mean value of \SI{402340}{cm^2/Vs} and a low variance of 18\%, indicating rather uniform heterostructure growth and device fabrication, on par with results reported in \cite{PaqueletWuetz2020} where the Hall bars were fabricated across a \SI{300}{mm} epitaxial wafer.
Since $n_\text{2D}$ is fixed in this definition of $\mu_0$, the variations observed in Fig.~\ref{fig3}(g)  must arise entirely from variations in $\rho_{xx}$.

It is common practice in the semiconductor industry to estimate the mobility (e.g., in an FET)  using the ``gradual channel’’ approximation~\cite{Sze2006}, leading to the definition $\tilde{\mu} \approx a(dI_\text{SD}/dV_\text{top})/(C'V_\text{SD})$, where $V_\text{SD}$ is the voltage dropped between the source and drain of the Hall bar.
We can check this approximation by extracting the average slopes $\overline{dI_\text{SD}/dV_\text{top}}$ from voltage sweeps like those shown in Fig.~\ref{fig3}(e), in the voltage range $\SI{20}{mV} < V_\text{top} < \SI{70}{mV}$, which is above the turn-on voltage.
Histogram results for $\tilde{\mu}$ are shown in Fig.~\ref{fig3}(h) (vermilion), indicating a narrow distribution of mobility estimates.
However, we also plot in blue the lower bounds on the mobility $\mu_\text{min}$, taken from sweeps like the one shown in Fig.~\ref{fig3}(c). 
This comparison indicates that $\tilde \mu$ typically underestimates the mobility, since the results are mostly smaller than $\mu_\text{min}$.
This is easily understood because the derivation of $\tilde\mu$ does not account for any dependence of $\mu$ on $V_\text{top}$, which is clearly inconsistent with Fig.~\ref{fig3}(d).

Finally, we use the Hall bar data to explore device properties across the sample.
Here again we make use of the mobility, which can reveal the dominant electron scattering mechanisms affecting $\rho_{xx}$. In particular, theory suggests that when mobility is dominated by effective 2D scattering from remote charged defects in the gate oxide, we should observe $\mu\propto n_{\rm{2D}}^{3/2}$.
Similarly, we should observe $\mu\propto n_{\rm{2D}}^{1/2}$ when the mobility is dominated by effective 3D scattering from background charges inside the quantum well~\cite{Monroe1993}. 
We expect the former to be valid in the low-density regime, while at higher densities, the latter relation should become dominant due to 2DEG screening of defects in the oxide~\cite{Mi2015}.
We test these theories by looking for the predicted slopes in the low-density (orange line) and high-density regimes (blue line), as shown in Fig.~\ref{fig3}(c).
These results can be further related to the 2D and 3D defect densities, $N_\text{2D}$ and $N_\text{3D}$, using the expressions given in \cite{Monroe1993}. 
The resulting estimates for defect densities are reported in the histogram data of Figs.~\ref{fig3}(i) and \ref{fig3}(j).

The disorder landscape in the quantum well can also be characterized using percolation theory, which describes the minimum density $n_\text{2D}$ needed to overcome disorder.
According to this theory~\cite{DasSarma2005}, conductivity data like Fig.~\ref{fig3}(b) should follow the scaling behavior $\sigma_{xx}\propto (n_{\rm{2D}} - n_p)^p$, where $n_p$ is the percolation density and $p$ is the percolation exponent.
We extract the percolation density in two ways in Fig.~\ref{fig3}(b).
In the first case, we allow both $p$ and $n_p$ to be fitting parameters (orange curve).
In the second case, we fix the theoretically expected exponent for a 2D system, $p = 4/3$, and only allow $n_p$ to be a fitting parameter (blue curve)~\cite{DasSarma2005,Tracy2009}. 
Note that near the percolation onset, the density is so low that the four-probe lock-in technique cannot accurately determine $n_\text{2D}$; in this regime, $n_\text{2D}$ values are therefore extrapolated from the formula $C'=n_\text{2D}/V_\text{top}+c_0$, using the values of $C'$ and $c_0$ determined previously.
Histograms of the percolation results are shown in Figs.~\ref{fig3}(k) and \ref{fig3}(l).
Figure~\ref{fig3}(l) indicates an average percolation exponent of $\bar{p} = 1.31$, with a relative standard deviation of 4\%, which is in excellent agreement with 2D percolation theory. 
We might expect the distribution of $n_p$ results in Fig.~\ref{fig3}(k) to track the distribution of $V_\text{T}$ in Fig.~\ref{fig3}(f), and indeed, we observe relative variations of 15\% ($p$ fixed) and 16\% ($p$ free) for the percolation density, compared to the relative variation of 18\% for $V_\text{T}$.

\section{Quantum Hall Effect}

\begin{figure} 
    \includegraphics[width = 0.7\columnwidth]{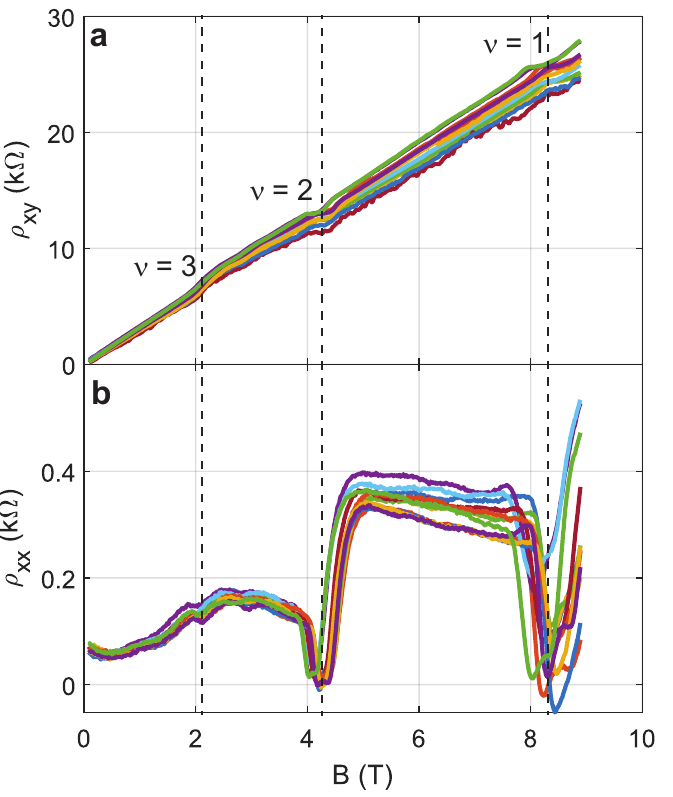}
    \caption{(a) Transverse ($\rho_{xy}$) and (b) longitudinal ($\rho_{xx}$) Hall resistivity scans as a function of magnetic field.
    Here, $\rho_{xy}$ exhibits quantized plateaus and $\rho_{xx}$ exhibits Shubnikov-de Haas oscillations, as expected for the integer quantum Hall effect.}
    \label{fig4}
\end{figure}

We now demonstrate the operation of the on-chip classical circuitry in the regime where the Hall bars studied are in the the integer quantum Hall regime~\cite{Girvin1990}.
In Hall bars, this effect manifests as quantized plateaus in $\rho_{xy}$ and Shubnikov-de Haas oscillations in $\rho_{xx}$, as observed in Fig.~\ref{fig4}.Here, the integer filling factors $\nu=1$, 2, and 3 are indicated by vertical dashed lines. 
The carrier density extracted from the slope of $\rho_{xy}$ vs $B$ in the low-field, classical Hall regime, is given by $n_{\rm{2D}} = \SI{2.06e11}{cm^{-2}}$.
The relation $\rho_{xy}=h/(\nu e^2)$, which is valid for integer $\nu$ values, also allows us to determine $n_\text{2D}$ in the quantum regime.
Using the minima of $\rho_{xx}$ to identify the locations of these integer values, we obtain $n_\text{2D}=\SI{2.00e11}{cm^{-2}}$ and \SI{2.02e11}{cm^{-2}} for the cases $\nu=1$ and 2, respectively.
This excellent agreement with the classical estimate suggests that no spurious conduction paths are present in our devices. 
In Fig.~\ref{fig4}(b), we note that these Shubnikov-de Haas minima can exhibit negative resistivity, which we argue in Supplementary Sec.~S4 can be explained as the Hall effect occurring in the FET switches.

Observing Shubnikov-de Haas oscillations represents an important milestone for on-chip multiplexing:
while previous multiplexers resided off-chip (e.g., in~\cite{PaqueletWuetz2020}) and were oriented perpendicular to the Hall bar to avoid spurious quantization effects, our work shows that measurements can still be performed when the classical switches are exposed to the same magnetic field and field orientation as the quantum devices, demonstrating their versatility. 
As explained in Supplementary Sec.~S4, operating the FETs well above their threshold voltage is key for suppressing the quantum Hall effect in these switches. 
We expect this to be an important consideration in future multiplexer designs, when it is necessary to operate quantum devices at high magnetic fields. 

\section{Multiplexed Readout}

\begin{figure*}
    \includegraphics[width = 1.95\columnwidth]{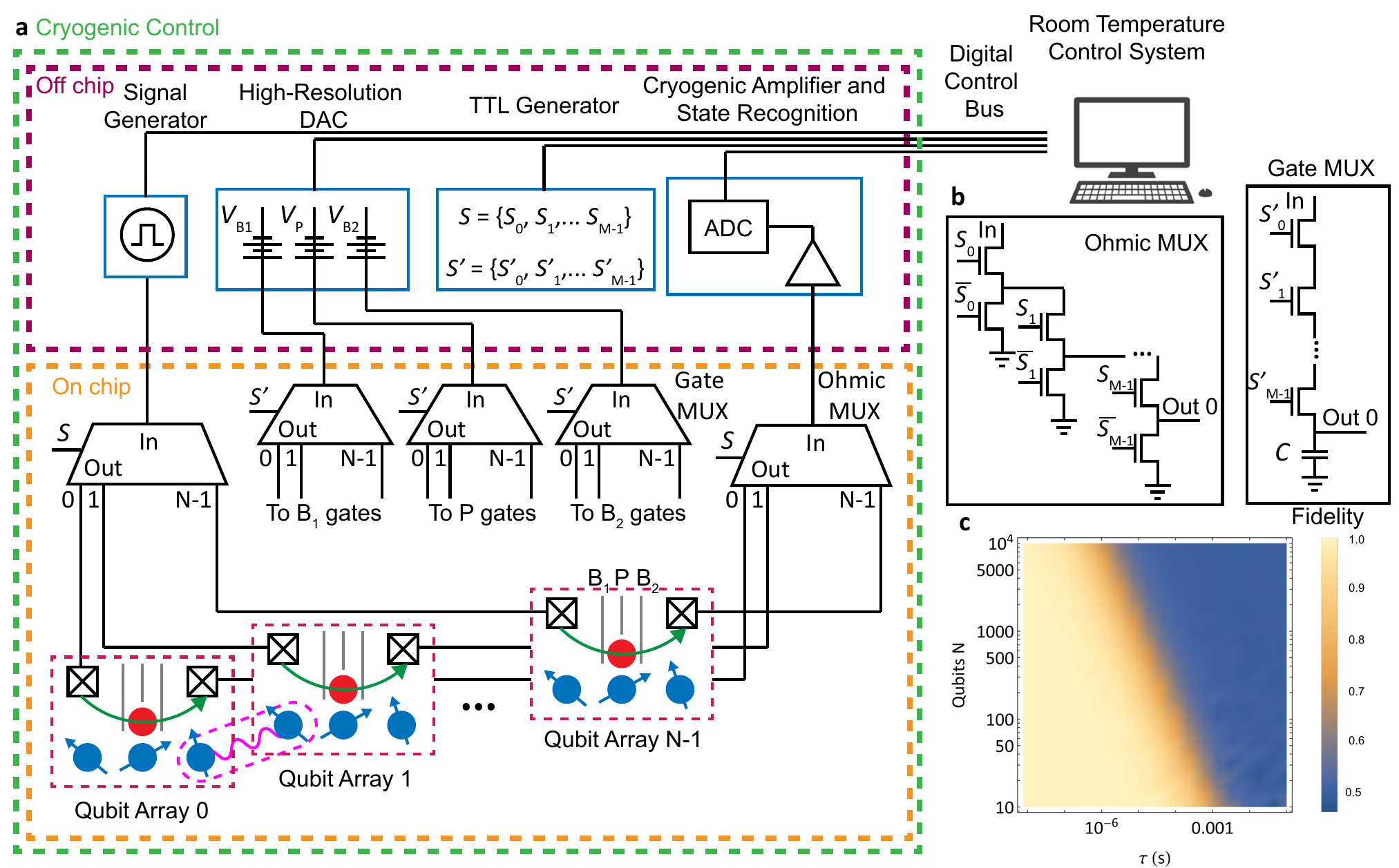}
    \caption{Scalable design for multiplexed readout of quantum-dot qubits.
    (a) Circuit schematic for connecting off-chip classical electronics (crimson box) with on-chip multiplexers (orange box), to alleviate wiring demands on the qubit arrays. 
    A charge-sensor dot (red) is positioned nearby qubits (blue) in a given array, with the sensor current (green arrow) flowing between ohmic contacts.
    Potential quantum couplers between arrays are shown as magenta connections.
    The ohmic contacts are addressed by the binary labels $S$, while the top gates (labeled B$_1$, P, B$_2$) are addressed by the binary labels $S'$. 
    The entire cryogenically controlled system (green box) is connected to a room-temperature control system via a digital control bus. 
    (b) FET schematics for multiplexing one ohmic contact (Qubit Array 0) and one top gate (also Qubit Array 0), for the addresses $S=S'=00\cdots 0$.  
    (c) Simulated average readout fidelity of an $N$-qubit array (assuming one qubit and one charge sensor per array) with a multiplexer rise time of $\tau$. 
    }
    \label{fig5}
\end{figure*}

Based on these results, we now present a scheme for multiplexing readout in gate-defined quantum-dot qubits, as illustrated in Fig.~\ref{fig5}(a).
In the figure, each of the $N$ arrays forms a construction zone for hosting quantum-dot qubits (blue) in close proximity to charge sensors (red), where the current through the charge sensor (green arrow) is used to measure the state of the qubits. 
The ohmic contacts on each array are multiplexed by on-chip FET switch circuits, as described above.
The input and output multiplexers route off-chip signal generators, cryogenic amplifiers, and analog-to-digital converters (ADCs) to and from the charge sensors.
These ohmic multiplexers are controlled by the binary address $S = \{S_0, S_1, ... S_{\mathrm{M-1}}\}$, where $M = \log_2(N)$ reflects the logarithmic scaling of the external lines. 
Our approach therefore significantly reduces the number of bulky and power-hungry cryogenic amplifiers~\cite{Blumoff2022,Mills2022} that can compromise available space and cooling power in a dilution refrigerator.

The plunger gates defining the quantum dots and tunnel barriers are also multiplexed by on-chip FETs, with a separate binary address $S'$.
However, in this proposed scheme for these gates we adopt the scheme described in \cite{Xu2020}, where the gate voltages are locked into place by charging an on-chip capacitor, labeled $C$ in Fig.~\ref{fig5}(b).
Since the FET is not used to flow currents to the top gates, we may use a simpler, non-complementary circuit here, like the one shown on the right-hand side of Fig.~\ref{fig5}(b), where the multiplexers route high-resolution digital-to-analog converter (DAC) voltage sources to the gate electrodes. 
Finally, we note that qubits in different arrays can be connected with long-range quantum couplers~\cite{Samkharadze2018,Mi2018, Mills2019, Holman2021, Seidler2022, Zwerver2023} (magenta connections), to provide space for on-chip multiplexers and other control wiring~\cite{Vandersypen2017}.

To assess the viability of such multiplexed readout, we consider a worst-case scenario in which every qubit array is queried, repeatedly (for example, in the context of quantum error correction).
In such a protocol, a qubit is first pulsed to its readout window. 
For simplicity, we have assumed one sensor per construction zone, although more sophisticated geometries could also be of interest.
Next, the ohmic multiplexer routes a current to the appropriate charge sensor, and then away to an off-chip cryogenic amplifier for digital state recognition.
When correctly tuned, the output signal should reflect the state of the qubit.
High readout fidelity requires a sufficiently long current integration time $t_{\mathrm{int}}$ to allow discrimination between the qubit states. We assume readout is performed by Pauli spin blockade~\cite{Johnson2005}, and we assume a readout-state lifetime of $T_1=10$~ms~\cite{Blumoff2022}.
The lifetime of the readout qubits $T_1$ should therefore be longer than the time needed to measure the whole $N$-qubit array in series, $T_{\mathrm{meas}} = Nt_{\mathrm{int}}$. 
Critical to this procedure is the readout signal rise time $\tau$, determined by the resistances and capacitances in the multiplexer and readout circuits. In our implementation, we measured $\tau =\SI{176}{\micro s}$, as described in Supplemental Sec.~3.  This time could be made faster by incorporating high-speed SiGe MODFET structures~\cite{Adesida1997}, or by reducing the parasitic capacitance between the routing layers. Since the readout signal can only change as fast as the rise time $\tau$, we adopt a single-qubit measurement time of $t_{\mathrm{int}} = 2\tau$, so that the readout circuitry reflects the state of the qubit without too much distortion from transients. Since gate pulses are typically much shorter than other readout time scales, we take these pulses to be instantaneous. Additional details of these simulations can be found in the Methods Section.

We plot in Fig.~\ref{fig5}(c) the readout fidelity predicted in this simulation. The results of the simulation provide design constraints for $\tau$ in the multiplexer. As the rise time $\tau$ goes down, more qubits can be measured in series for a fixed time $T_{\rm{meas}}$, assuming the integration time for each qubit measurement can be made small. In contrast, as $T_{\rm{meas}}$ increases, an increasing fraction of qubits will relax while in queue for readout for a fixed $T_1$ relaxation time. The results of Fig.~\ref{fig5}(c) show that multiplexing can indeed reduce the number of gate lines required to perform readout, a capability that will be increasingly important as the size of semiconductor quantum processors continues to grow.
 
\section{Summary}

We have fabricated and studied a multiplexing platform capable of controlling 16 quantum devices in Si/SiGe. 
Our method utilizes on-chip switching electronics that enable rapid, in-situ characterization of a large array of devices, resulting in a nearly ten-fold reduction in electrical interconnects. 
The on-chip switches allow us to quickly acquire a statistical set of data for assessing device uniformity, using an array of Hall bars fabricated by electron-beam lithography. We focus here on key transport properties that are commonly employed as indicators for successful operation of quantum-dot qubits, including threshold voltage, gate capacitance, percolation density, electron mobility in gate-defined devices, and defect densities.  By observing Shubnikov-de Haas oscillations, our technology bridges classical and quantum circuits and paves the way for integrating on-chip multiplexing electronics in large-scale quantum computers in the solid state.

\section*{Acknowledgements}

We thank HRL Laboratories, LLC for support and L.F. Edge for providing the Si/SiGe heterostructure used in this work. We thank Lisa Tracy for useful discussion. Research was sponsored in part by the Army Research Office (ARO) under Awards Number W911NF-17-1-0274 and W911NF-23-1-0110. The views and conclusions contained in this document are those of the authors and should not be interpreted as representing the official policies, either expressed or implied, of the Army Research Office (ARO), or the U.S. Government. The U.S. Government is authorized to reproduce and distribute reprints for Government purposes notwithstanding any copyright notation herein. Sandia National Laboratories is a multimission laboratory managed and operated by National Technology and Engineering Solutions of Sandia, LLC, a wholly owned subsidiary of Honeywell International Inc., for the U.S. Department of Energy’s National Nuclear Security Administration under contract DE-NA0003525.

\section*{Methods}

\noindent\textbf{Heterostructure and Device Fabrication.}
Device fabrication begins with an undoped Si/SiGe heterostructure formed on a low-resistivity Si wafer. The heterostructure is grown via chemical vapor deposition (CVD) and begins with a polished \SI{3}{\micro\meter} linearly graded \sige relaxed buffer substrate, followed by a \SI{400}{nm} thick \sige layer, a \SI{10}{nm} pure-Si quantum well, an \SI{80}{nm} \sige spacer, and a \SI{2}{nm} protective Si cap. A \SI{20}{nm} padding of \SiOx (oxide) is grown by CVD to protect the substrate from resist liftoff after ion implantation. 
The full heterostructure is illustrated in Fig.~\ref{fig2}(f). 

A CHF$_{3}$ plasma etch removes the quantum well everywhere except for sixteen mesas, with a target thickness of \SI{130}{nm}, which form the construction zones. This etch layer also forms Si channels for the field-effect-transistor (FET) switches. Ohmic contacts are formed using a $^{31}$P$^+$ ion implant, a forming-gas anneal, and a \SI{2}{nm}/\SI{50}{nm} Ti/Pd metallization. The padding oxide is then stripped and replaced with \SI{20}{nm} of \SiOx, which acts as a gate oxide for the FET switches and a field oxide for metal deposition steps that come later. 

Two Ti/Pd routing layers, \SI{2}{nm}/\SI{50}{nm} and \SI{10}{nm}/\SI{100}{nm}, with \SI{100}{nm} of isolation oxide deposited between the layers, are fabricated using standard optical lithography. Vias are etched for electrical contact between these layers. The gate oxide for the construction zone begins with a \SI{72}{\micro\meter}$\times$\SI{72}{\micro\meter} field-oxide etch, followed by \SI{10}{nm} of \HfOx deposited via atomic-layer deposition. The construction zone fabrication is completed with \SI{2}{nm}/\SI{20}{nm} Ti/Pd jumpers to alleviate step coverage for structures fabricated via electron-beam lithography. Six ion-implanted regions and fifteen palladium jumpers border the perimeter of the mesa.

The Hall bars measured in the experiments are formed using palladium top gates patterned by electron-beam lithography. 
Figure~\ref{fig2}(e) shows a micrograph of Device~0. 
The Hall bar top gates are tied together and controlled via a global voltage \Vtop. 
The full device shown in Fig.~\ref{fig2}(a), including all 16 Hall bars and multiplexing circuitry, requires a total of sixteen electrical connections at the bond pads: one top gate, six ohmics, eight switch control lines ($S_0$-$S_3$ and their complements $\bar S_0$-$\bar S_3$), and a ground line. 
This represents a nearly ten-fold reduction of control lines compared to sixteen separately wired Hall bars, which would require 112 electrical connections. 

In this work, we report on results obtained from 12 of the 16 Hall bars, since four of devices displayed flaws: 
devices $S = 0010$ and 1010 did not turn on in the range of \SI{0}{}-\SI{100}{mV}; device $S = 1100$ exhibited no transverse voltage; device $S=1111$ had a broken drain contact.

\noindent\textbf{Readout Fidelity Simulations.}
We simulate Pauli-spin-blockade readout, which makes use of the fact that the $\ket{S(0,2)}$ and $\ket{T(1,1)}$ states induce different responses from the charge sensor.
Here we assume that the (1,1) $\rightarrow$ (0,2) transition is instantaneous, and that the readout state has a typical singlet-triplet decay time of $T_1 = \SI{10}{ms}$ in Si/SiGe double dots~\cite{Blumoff2022}.
The last (i.e., the $N^\text{th}$) qubit in the serial readout scheme therefore has a probability of $e^{Nt_\text{int}/T_1}$ to relax to the $\ket{S(0,2)}$ before readout is complete. 

For the simulations, a string of $N$ bits is randomly generated, representing the final state to be read out after a gate operation. $N$ serial readout steps are then performed over a total measurement period of $Nt_{\rm{int}}$, simulating $N$ queries from the multiplexer in the proposed readout scheme. The readout fidelity is calculated by comparing each pulse in the string to a threshold value, then averaging over 50 randomly generated $N$-length strings. In the tan-colored region of Fig.~\ref{fig5}(c), the multiplexer speed is high enough for all $N$ qubits to be read out before the qubits relax. In contrast, in the blue region, the multiplexer is slow enough that the qubits relax before the serial measurement is complete. 

\bibliography{ref.bib}

\newpage
\vfil

\pagebreak
\renewcommand\thesection{S\arabic{section}}
\renewcommand\thefigure{S\arabic{figure}}
\renewcommand\theequation{S\arabic{equation}}
\setcounter{figure}{0}
\setcounter{equation}{0}
\setcounter{section}{0}
\onecolumngrid

\onecolumngrid

\vspace{0.5in}
\begin{center}
\textbf{\large Supplementary Information}
\end{center}

\section{Chip image and CAD design}
\begin{figure*}[b]
    \includegraphics[width = 7in]{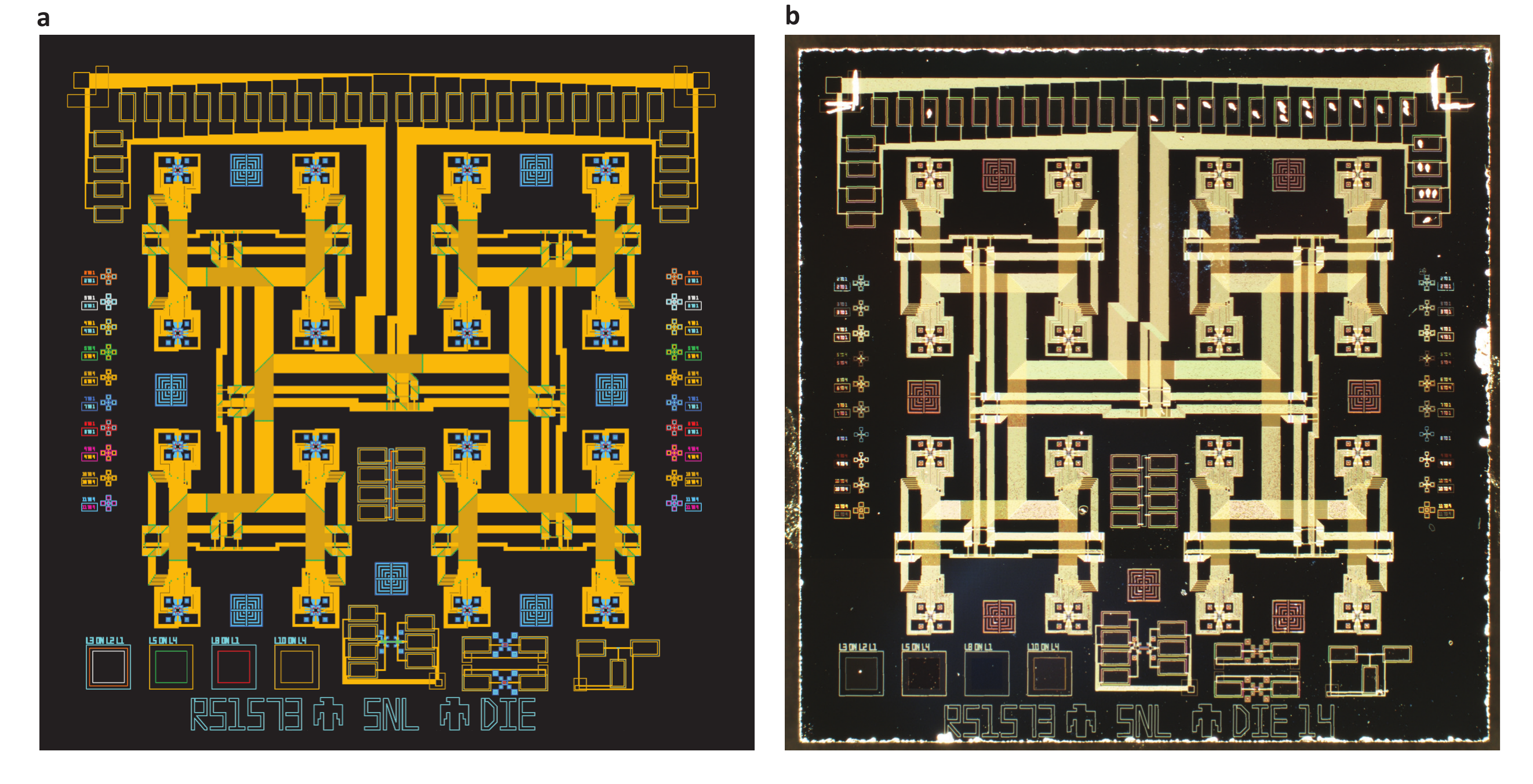}
    \caption{(a) CAD representation of the device. (b) Micrograph of a full chip, which is lithographically identical to the chip measured in this work.} 
    \label{sfig1}
\end{figure*}

Figure~\ref{sfig1}(a) shows the CAD layout of the chip used in our experiment. Figure~\ref{sfig1}(b) shows a lithographically patterned device, which we note is stitched together from micrographs of the 11.5x11.5mm$^2$ chip using two images taken from a Wild 420 Macroscope. 

\section{FET operation}

\begin{figure}
    \includegraphics[width = 6in]{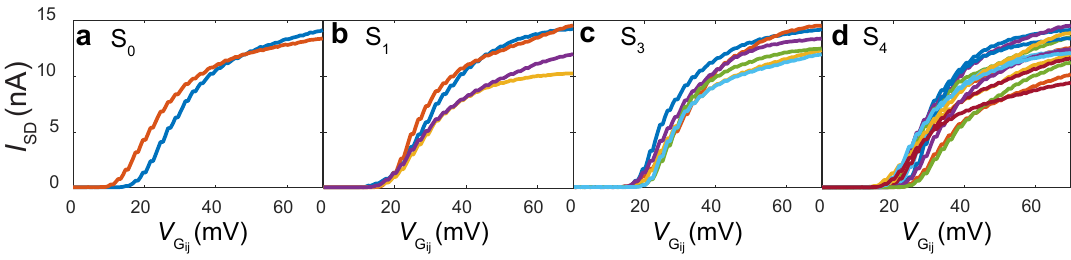}
    \caption{
    Pinch-off curves for the FET switches: (a) FETs for $S_0$ and $\bar S_0$; (b) FETs for $S_1$ and $\bar S_1$; (c) FETs for $S_2$ and $\bar S_2$; and (d) FETs for $S_3$ and $\bar S_3$. 
    Here, the bar notation refers to the complimentary FET gates in each switch. That is, when $S_0 = 0$, $V_{S_0} = V_{G_{21}} = V_{G_{12}}$ is high, and $V_{\bar S_0} = V_{G_{11}} = V_{G_{22}}$ is low. 
    }
    \label{sfig2}
\end{figure}

In this section we provide data to show that the multiplexer is functioning correctly. We assign each Hall bar a four-bit digital address $S = \{S_0, S_1, S_2, S_3\}$. As described in section~III, the binary value of $S_j$ represents the logical state of the switches in the array. In  Figs.~\ref{sfig2}(a)-\ref{sfig2}(d) we show pinch of curves for each FET in the switch array, confirming the correct current path to the Hall bar selected by each digital address. There are two current paths possible for the $S_0$ switch, four for the $S_1$ switch, and so on. The current paths for the Hall bars known to be not functional are omitted.  

When the FETs are turned on, we ensure that they are operated well above the threshold voltage ($\approx$ \SI{15}{mV}), at high accumulation density, to ensure minimal magnetoresistance when characterizing the quantum Hall effect in the devices under test (see also Sec.~S4, below). When the FETs are turned off, they are tuned to well below the threshold, to mitigate leakage-current paths between the switch array and ground.

\begin{figure*}
    \centering
    \includegraphics[width = 7in]{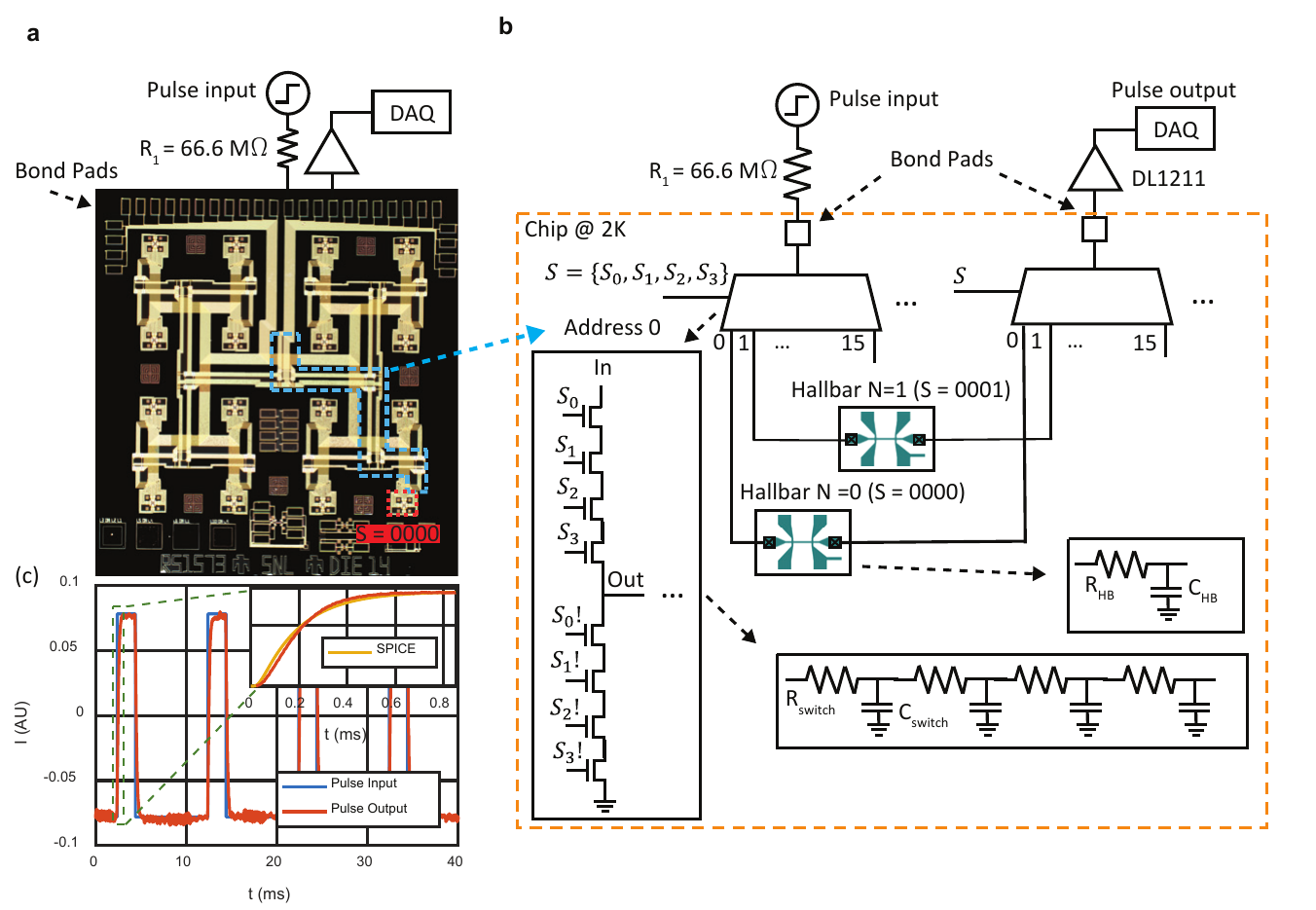}
    \caption{Measurement of multiplexer bandwidth. (a) Circuit connections to the multiplexing chip. 
    (b) Circuit schematic showing the elements inside the blue box indicated in (a).
    (c) Square current pulses of width \SI{2}{ms} are sent to the Hall bar device $S = 0000$, to measure the capacitance to ground.}
    \label{sfig4}
\end{figure*}

\section{Multiplexer bandwidth}

The capacitance of the multiplexer is determined by sending a current pulse to the Hall bar located in the $S = 0000$ construction zone, as indicated in Fig.~\ref{sfig4}(a), and measuring the output current. Here, the pulse is generated by a Tektronix 3011 AFG in series with the resistance $R_1 = \SI{66.6}{M\ohm}$ chosen to emulate a current source. As shown in Fig.~\ref{sfig4}(b), the current pulse arrives at a bond pad, propagates through the multiplexer via four FET switches (outlined in blue in Fig.~\ref{sfig4}(a)), arrives at the Hall bar, and exits via the multiplexer circuit to a room-temperature DL1211 current preamplifier and data acquisition unit (DAQ). The signal path is modeled with four FET switches, each with a thru resistance of $R_{\rm{switch}}$ and capacitance $C_{\rm{switch}}$. The latter accounts for both the capacitance of the FET (top gate to 2DEG) and the cross-capacitance from the densely packed and overlapping meandering wires. The Hall bars are modeled with source-drain resistance $R_{\rm{HB}}$ and capacitance $C_{\rm{HB}}$, where $C_{\rm{HB}}$ is simply the gate-to-2DEG capacitance. All other ohmics on the Hall bar are allowed to float. The output signal shows a clear RC rise time (\SI{176}{\micro s}) [Fig.~\ref{sfig4}(c) inset], resulting from a combination of resistances and capacitance in the circuit. We have verified that the room-temperature current amplifier does not limit the response of the input pulse. We then fit the rise-time curve using a spice simulation with parameters $R_{\rm{switch}} = \SI{10}{k\ohm}$, $R_{\rm{HB}} = \SI{2.4}{k\ohm}$, $C_{\rm{HB}} = \SI{3.3}{pF}$, to extract a multiplexer capacitance of $C_{\rm{switch}} = \SI{14.4}{nF}$. We note that this value is much larger than the gate-to-2DEG capacitance of the FETs in the switch, and we therefore attribute it to capacitance between routing layers. 

\begin{figure}
    \centering
    \includegraphics[width = 4in]{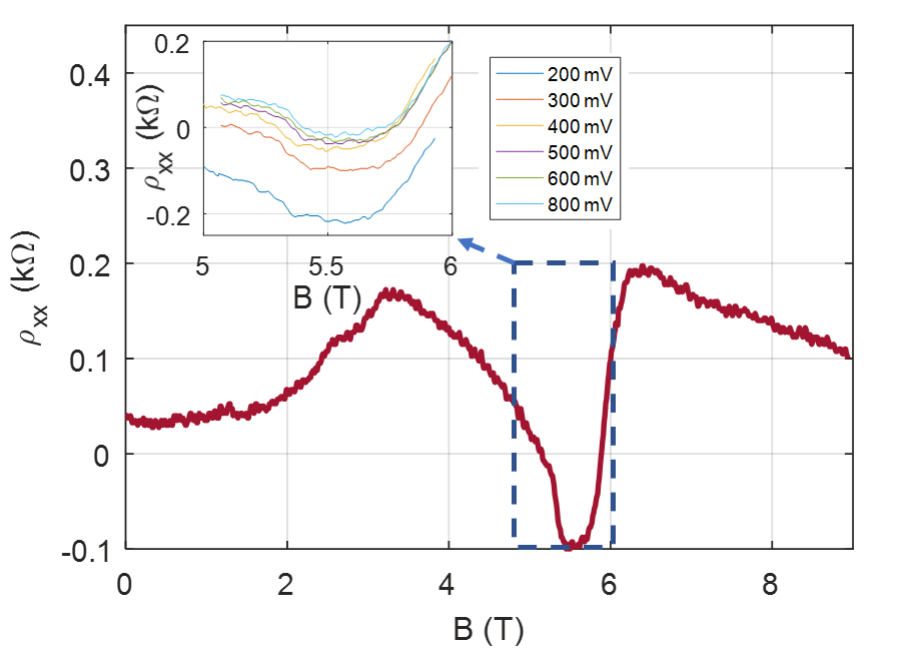}
    \caption{Longitudinal resistance $\rho_{xx}$ of the Hall bar labelled $S = 0000$, as a function of magnetic field. Inset: Longitudinal resistances obtained in the range $\SI{5}{} < B < \SI{6}{T}$, for several different switch FET gate voltages in the range $\SI{200} < V_{G_{12}} = V_{G_{21}} < \SI{800}{mV}$.}
    \label{sfig3}
\end{figure}

\section{Hall effect in the multiplexer}

The integrated FET switches are fabricated in the same plane as the Hall bars; therefore the multiplexing circuitry experiences the same magnetic field as the Hall bars. We observe a dependence of $\rho_{xx}$ on the switch FET gate voltage. Figure~\ref{sfig3} shows a high-field sweep of $\rho_{xx}$ for the $S = 0000$ Hall bar, which is typical for a quantum Hall effect experiment. Here, we used gate voltages $V_{G_{12}} = V_{G_{21}} = \SI{300}{mV}$ and $V_{G_{11}} = V_{G_{22}} =  \SI{-1.5}{V}$ to implement the appropriate digital address. We find that the region where $\rho_{xx}$ becomes negative (outlined by the dashed blue box) is suppressed as $V_{G_{12}}$ and $V_{G_{21}}$ becomes more positive (see inset). This suggests that $\rho_{xx}$ is affected by Hall effect voltages in the switches, when the 2DEG density under the FET gates is low. The behavior of $\rho_{xx}$ returns to its expected range ($\rho_{xx} > 0$) when the voltage on the FET gates are higher than \SI{600}{mV}. To avoid such behavior, the quantum Hall effect data presented in the main text is always obtained using gate voltages $V_{G_{12}} = V_{G_{21}} = \SI{700}{mV}$ and $V_{G_{11}} = V_{G_{22}} = \SI{-1.5}{V}$.

\end{document}